\documentclass[fleqn,usenatbib]{mnras}

\usepackage{layouts}
\usepackage{mathptmx}
\usepackage[T1]{fontenc}

\DeclareRobustCommand{\VAN}[3]{#2}
\let\VANthebibliography\thebibliography
\def\thebibliography{\DeclareRobustCommand{\VAN}[3]{##3}\VANthebibliography}

\usepackage{graphicx}	
\graphicspath{{figures/}}
\usepackage{amsmath}	
\usepackage{amssymb}	
\usepackage{amstext}
\usepackage{siunitx}
\usepackage{mathtools}
\usepackage{color}
\usepackage{twoopt}
\usepackage{hyperref} 
\usepackage{afterpage}
\usepackage{silence}
\usepackage{newtxtext,newtxmath}

\title[Observability of Transition Discs]{Observability of Photoevaporation Signatures in the Dust Continuum Emission of Transition Discs}

\author[Picogna et al. 2023]{
Giovanni Picogna$^{1}$,\thanks{E-mail: picogna@usm.lmu.de}
Carolina Sch{{\"a}}fer$^{1}$,
Barbara Ercolano$^{1}$,
Christian Rab$^{1}$,
Rafael Franz$^{1}$,
and Mat\'ias G\'arate$^{2}$
\\
$^{1}$Universit\"ats-Sternwarte, Ludwig-Maximilians-Universit\"at M\"unchen, Scheinerstr. 1, 81679 M\"unchen, Germany\\
$^{2}$Max-Planck-Institut f\"ur Astronomie, K\"onigstuhl 17, 69117, Heidelberg, Germany
}

\date{Accepted 2023 May 9. Received 2023 May 8; in original form 2023 January 12}

\pubyear{2023}

\begin{document}
\label{firstpage}
\pagerange{\pageref{firstpage}--\pageref{lastpage}}
\maketitle

\begin{abstract}
Photoevaporative disc winds play a key role in our understanding of circumstellar disc evolution, especially in the final stages, and they might affect the planet formation process and the final location of planets. The study of transition discs (i.e. discs with a central dust cavity) is central for our understanding of the photoevaporation process and disc dispersal. However, we need to distinguish cavities created by photoevaporation from those created by giant planets. Theoretical models are necessary to identify possible observational signatures of the two different processes, and models to find the differences between the two processes are still lacking. In this paper we study a sample of transition discs obtained from radiation-hydrodynamic simulations of internally photoevaporated discs, and focus on the dust dynamics relevant for current ALMA observations. We then compared our results with gaps opened by super Earths/giant planets, finding that the photoevaporated cavity steepness depends mildly on gap size, and it is similar to that of a \SI{1}{M_J} mass planet. However, the dust density drops less rapidly inside the photoevaporated cavity compared to the planetary case due to the less efficient dust filtering. This effect is visible in the resulting spectral index, which shows a larger spectral index at the cavity edge and a shallower increase inside it with respect to the planetary case. The combination of cavity steepness and spectral index might reveal the true nature of transition discs.
\end{abstract}




\section{Introduction}

Transition discs are defined as circumstellar discs with a dust inner cavity. Historically, this class of discs has been detected by studying their spectral energy distribution (SED), which showed a dip in the near-infrared part of the spectra, explainable by a lack of warm dust emitting close to the central star \citep{Strom1989}. With the advent of high-resolution imaging of protoplanetary discs in the (sub)millimeter continuum emission, especially thanks to the Atacama Large Millimetre Array (ALMA), we are now able to probe directly the dust content in the inner regions of planet forming discs, and detect potential cavities.

There are two leading theories to explain this class of discs.
The first is the theory of disc photoevaporation which has its foundation in the seminal work by \cite{Hollenbach1994} and explains the observed fraction of transition discs ($\sim 10\%$ of the entire disc population from SED studies) in a "two-time-scale" scenario \citep{Clarke2001}. The X-ray and UV radiation from the central star drives a constant thermal wind from the cicumstellar disc that, where and when eventually the accretion rate drops below the wind mass-loss rate, creates a gapped disc that is dispersed on a short timescale, which is roughly $1/10$-th of the total disc life-time \citep{Alexander2014,Ercolano2017}.
The second theory of transition disc formation invokes the presence of super-Earth/giant planets that can halt the flux of dust from the outer disc by creating a pressure bump outside their location, and lead to a depleted inner dust disc. This second scenario would explain the observed fraction of transition discs as a probability of $10\%$ of forming a giant planet on a wide orbit \citep{Duffell2015,Dong2016}

Recent ALMA observations have questioned this fraction of transition discs by observing several unresolved dust cavities where the SED showed a dip in its near-infrared emission, and vice versa, showing dust cavities in discs where the SED didn't show any characteristic features \citep{Andrews2018,Kurtovic2021,vanderMarel2022}.
The fraction of giant planet on wide orbits seems to be also much smaller than the $10 \%$ value necessary to explain the SED population \citep{Nielsen2019,Vigan2021}.

There have been several theoretical studies to identify the region of the parameter space that the two different models can explain, with disc photoevaporation able to create small cavities with low accretion rate onto the central star, and giant planets on the other side of the spectra, creating mainly cavities where the accretion rate onto the central star is still unperturbed \citep{Ercolano2017}.
However, recent models have questioned this view pushing the region of the parameter space explainable by disc photoevaporation both to larger cavity sizes and to higher accretion rates \citep{Picogna2019,Ercolano2018,Garate2021}.

The best way to distinguish between the two mechanisms, if they don't occur in combination, is possibly trough dust observations. \citet{Franz2022} showed that dust entrained in photoevaporative winds might leave observable features that can be detected in the near future.
Another avenue is to understand the effect of a dust cavity opened by photoevaporation on bigger dust particles settled towards the disc mid-plane and directly observable with modern facilities like ALMA.
In this work we study how disc photoevaporation affects the dust distribution close to the mid-plane and the cavity edge, and compare the results with literature in order to understand the characteristic fingerprints of photoevaporation and how to distinguish them from a planetary scenario.

In Section~\ref{sec:methods} we lay down the methods adopted to perform 2D dusty hydrodynamic simulations of transition discs photoevaporated from the central star, and the post-processing with Monte Carlo radiative transfer simulations. In Section~\ref{sec:results} we present the main results of our work, which we then compare with previous literature to understand its characteristic features in Section~\ref{sec:discussion}.
Finally, in Section~\ref{sec:conclusions} we present the main conclusions.

\section{Methods}\label{sec:methods}

    In this section we present our radiation-hydrodynamic model for transition discs which includes X-ray photoevaporation from the central star and dust dynamics. When an equilibrium configuration is reached, we extract the radial and vertical dust distribution and derive synthetic images from detailed radiative transfer calculations.

    \subsection{Hydrodynamics}\label{sec:hydro}

        We used the hydro-code {\sc pluto} \citep{Mignone2007} with the modifications of \citet{Picogna2019} to account for the stellar X(EUV) irradiation. 
        We adopted a spherical coordinate system centred onto the star.
        In this reference frame we can compute, without increasing the computational load, the column density from the central star at each cell.
        Using the column density and the disc local properties we determine the temperature in the bulk of the disc and in the disc atmosphere as explained in Sec.~\ref{sec:mocassin}.
        The initial density structure was taken from the D’Alessio Irradiated Accretion Disc (\textsc{diad}) radiative transfer models \citep{DAlessio1998,DAlessio1999,DAlessio2001,DAlessio2005,DAlessio2006} code that provide also the gas temperature in the bulk of the disc that best fit the median spectral energy distribution (SED) in Taurus. For more details, the reader is referred to \citet{Picogna2019}. A cavity was then created by an exponential cut-off of the density
        \begin{equation}
          \rho = \rho_0 \cdot \exp{\left[\frac{2\pi(R-R_\textrm{cav})}{\Delta_\textrm{gap}}\right]}\, \textrm{for}\, R \leq R_\textrm{cav},
        \end{equation}
        where $\rho_0 = \rho(R_\mathrm{cav})$ is the initial density at the cavity location, $R$ is the cylindrical radius, $\Delta_\textrm{gap} = 1$ au is the width of the cavity wall, and $R_\textrm{cav}$ is the cavity radius.
        We explore three different transition disc setups with cavity sizes of $10$, $20$, and $30$ au respectively.
        The main physical properties of the numerical simulations are reported in Table~\ref{tab:PlutoPars}.
           
        \begin{table}
        \centering
        \begin{tabular}{c c c c c c c}
         \hline
         $L_x$ & $M_\star$ & $R_\textrm{cav}$ & $\alpha$ & $s$ & $\rho_s$\\ 
         \hline
         erg/s & $M_\odot$ & au & & cm & g cm$^3$\\
         \hline
         $2\cdot 10^{30}$ & $0.7$ & $10, 20, 30$ & $0.001$ & 0.01,0.1,1,10 & 1 \\ 
        \hline
        \end{tabular}
        \caption{\label{tab:PlutoPars} Main properties of the parameter space studied.}
        \end{table}

        The grid radial coordinate $R$ extends from just inside the cavity location (at 5, 10, and \SI{15}{au} for increasing gas cavity sizes) to allow an increased resolution of the cavity edge and at the same time save the computational time needed to model the empty region inside it. The outer radial boundary is placed at \SI{1000}{au}, with an outer region from \SI{800}{au} where the hydrodynamics is not evolved, in order to explore the entire extent of the disc and avoid numerical reflections that would alter the wind diagnostics \citep{Picogna2019}. The radial grid is divided into 300 log-spaced cells, and open boundary conditions are applied at the inner and outer edges.
        In the polar direction, the grid ranges from the polar axis ($\theta = 0.005$) to the mid-plane with $160$ logarithmically spaced cells, which gives a higher resolution close to the disc mid-plane where the dust dynamics is taking place. The boundary condition is reflective both at the mid-plane and close to the polar axis. The grid is 2-dimensional, since the problem is symmetric in the azimuthal direction.
        
    \subsection{Thermal Calculations}\label{sec:mocassin}

        The gas temperature in the disc atmosphere is determined by a parameterisation based on detailed multi-frequency gas photoionisation and dust radiative transfer calculations performed with the {\sc mocassin} code \citep{MOCASSIN1,MOCASSIN2,MOCASSIN3} for typical protoplanetary disc conditions, irradiated by a realistic XEUV spectrum presented in \citet{Ercolano2008,Ercolano2009}.
        The resulting temperature is dependent on the column density to the central star $N$ and the ionisation parameter $\xi=L_x/nr^2$, where $L_x$ is the stellar X-ray luminosity, $n$ the local gas particle density, and $r$ the distance from the star, and it is tabulated for fast calculation.
        This temperature prescription is valid up to the maximum penetration depth of X-rays ($N_X = 2\cdot10^{22}$ pp/cm$^2$, \cite{Ercolano2009}).
        For larger column densities we assume perfect dust and gas thermal coupling and adopt the dust temperatures from hydrostatic disc models obtained via the \textsc{diad} radiative transfer models.
        The gas (and dust) temperature is thus tabulated and locally isothermal. We do not include effects like adiabatic cooling, viscous heating, dust-gas thermal interactions.
       
    \subsection{Dust dynamics}\label{sec:dust}
    
        The simulation was initialized only with the gas component for a few tens of orbits (at 10 au) to allow for the disc to reach a vertically hydrostatic equilibrium before the dust was inserted.
        We modelled the dust component of the disc with $4\cdot10^5$ Lagrangian particles equally distributed in four size bins (relevant to ALMA dust continuum observations) with a radius $s$ ranging from \SI{100}{\mu m} to \SI{10}{cm} (see Table~\ref{tab:PlutoPars}).
        From a dynamical point of view, particles smaller than $100 \, \mu$m are almost completely coupled to the gas. Decreasing the minimum size has been found to have a very limited impact also on our synthetic maps for the wavelength considered, so we restricted our analysis to this range.
        The dust particles were placed with an initial scale height \citep{Youdin2007}
        \begin{equation}\label{eq:scale-height}
            \frac{H_p}{H} = \sqrt{\frac{\alpha}{\alpha + St}}\,,
        \end{equation}
        where we assumed an isotropic turbulence ($\alpha_z = \alpha_R = \alpha = 0.001$), $H$ is the local gas scale height, $St = t_s\Omega_K$ is the particle Stokes number, $t_s$ is the particle stopping time, which in the Epstein regime (for dust sizes much smaller than the mean free path of gas molecules) takes the form
        \begin{equation}\label{eq:Stokes_number}
            t_s = \frac{s\rho_s}{\rho_g\bar{v}_{th}}\,,
        \end{equation}
        $\rho_s$ is the dust internal density, $\rho_g$ the local gas density, $s$ the dust size, and $\bar{v}_{th}$ the local gas thermal velocity. In this way, the initial dust configuration is close to its equilibrium and we limit vertical oscillations.
        The Stokes number (or analogously the stopping time) is a measure of the dust-gas coupling. Dust with a small Stokes number is tightly coupled with the gas, following closely its distribution, while for increasing Stokes number the dust starts to decouple from the gas, not feeling the pressure support, drifting towards the central star and settling on the disc midplane \citep{Whipple1972,Weidenschilling1977}. The dust with Stokes number of order unity are those that evolve more rapidly with respect to the gas.
        The four dust sizes modelled correspond to minimum Stokes numbers of $1.4\cdot10^{-3}, 1.2\cdot10^{-2}, 9.6\cdot10^{-2}, 0.98$ respectively (calculated at the disc midplane for the 10 au transition disc case).
        Any dust drifting through the radial boundaries of the simulation was reinserted at the outer boundary to keep the number of dust particles constant, while reflecting conditions were applied for the azimuthal boundaries, similarly as for the gas component.
        The particles were evolved using a leap-frog integrator.
        To prevent the dust to settle vertically and keep the initial equilibrium scale height given by equation~\ref{eq:scale-height}, we apply random turbulent kicks to the dust particles proportional to the level of turbulence $\alpha$ adopted (see Table~\ref{tab:PlutoPars}). For more details on the numerical implementation, the reader is referred to \cite{Picogna2018}. We ran the hydrodynamical simulations together with the dust evolution for $\sim15$ kyr, at which point the dust distribution had reached a stable configuration for all modelled dust sizes.
        
     \subsection{Radiative transfer} \label{sec:radiative transfer}
     
         The results of the hydrodynamical simulations were then further processed with the radiative transfer code \textsc{radmc3d} \citep{radmc3d} to obtain improved dust temperatures and dust continuum emission maps at $0.85$ and $3$ mm, which can be observed in Band 7 and 3 by ALMA. The grid was first extended in the azimuthal direction with 600 cells, using a modified version of \textsc{fargo2radmc3d} code \citep{Baruteau2019}. The dust size distribution $n$ used for the radiative transfer calculation is a power law function of particle size $a$, $dn/da \propto a^{-3.5}$ between the minimum and maximum sizes modelled. We then compute the dust surface density of each size bin from its spatial distribution as:
         \begin{equation}
            \sigma_{i,\mathrm{dust}}(r,\theta) = \frac{N_i(r,\theta)}{A(r)} \frac{M_{i,\mathrm{dust}}}{\sum_{r,\theta}N_i(r,\theta)}\,,
         \end{equation}

         where $N_i$ is the number of dust particles per bin size in each grid cell of the simulation, $A$ is the surface area of each grid cell (seen face-on), and 
         \begin{equation}
            M_{i,\mathrm{dust}} = \xi M_\mathrm{gas} \frac{a_i^{4-p}}{\sum_i a_i^{4-p}}
         \end{equation}
         is the dust mass per bin size, with $p=3.5$ the power-law exponent of the dust size distribution, $\xi=0.01$ the dust-to-gas mass ratio, and $M_\mathrm{gas}$ is the total gas mass in the simulation.

         For the optical depth calculation we assumed a dust opacity of compact spherical grains from Mie theory and adopt optical constants of astronomical silicates \citep{Draine2003}.

         We place the disc at a distance of \SI{140}{pc}, which is typical of close star-forming regions.
         The temperature and radius of the star were set to be $T_\mathrm{eff}=5000 K$ and $R_*=2.5 R_\odot$, consistent with the X-ray luminosity adopted in the hydrodynamic model.
         We chose a logarithmically-spaced wavelength grid $10^{-1} \leq \lambda [\mu m] \leq 10^4$, $10^8$ photon packages for ray-tracing, creating a $2048\times2048$ pixel flux map.
         Finally, we create mock observations by convolving the intensity maps of the modelled images with ALMA beam sizes for Band 7 (\SI{0.85}{mm}) and Band 3 (\SI{3}{mm}). Specifically, we took the same set-up used in \cite{Nazari_2019} and adopted the configurations C43-8 and C43-5 for Band 7 giving a beam size of $0.034 \arcsec \times 0.031 \arcsec$, and C43-9, C43-6 for Band 3 with a beam size of $0.067 \arcsec \times 0.056 \arcsec$.

\section{Results}\label{sec:results}

    \subsection{Hydrodynamics} \label{sec:results-gas}

\begin{figure*}
            \centering
            \includegraphics[width=\textwidth]{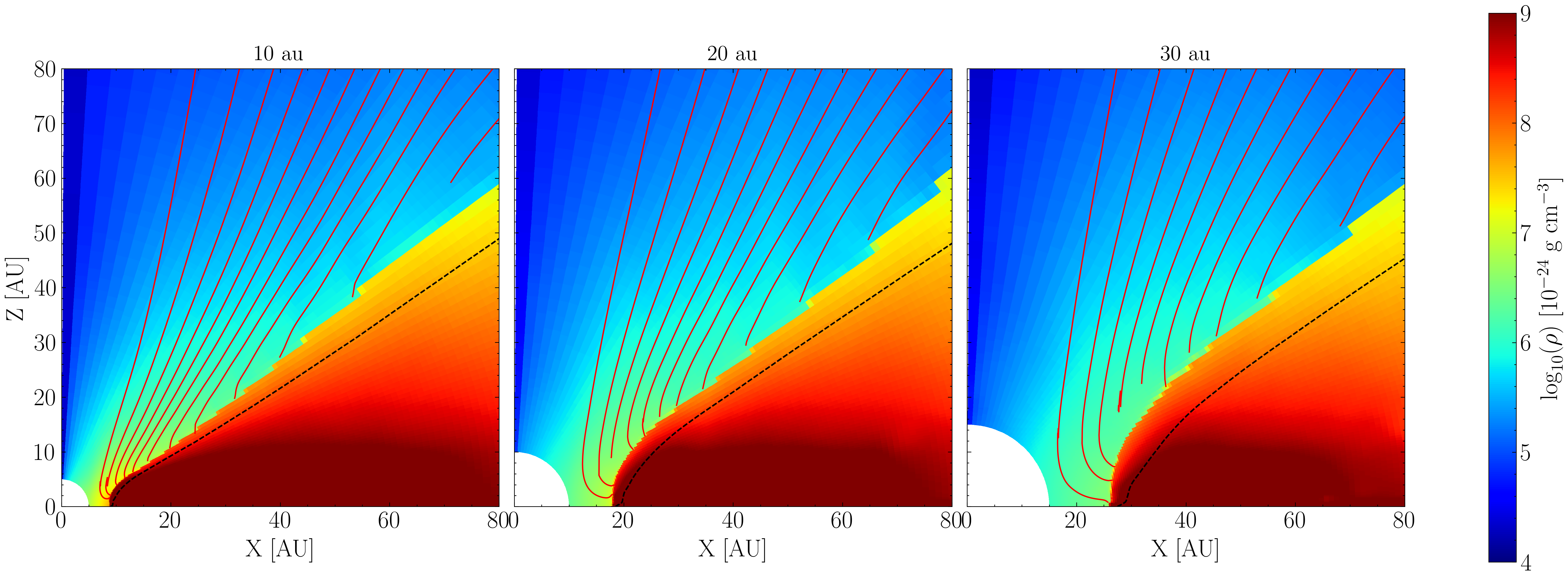}
            \caption{Gas density distributions for the different runs in the inner \SI{80}{au}. The black dashed line show the maximum penetration depth of X-rays ($N_X = 2\cdot 10^{22}$ pp cm$^{-2}$), while the red lines show gas streamlines in the wind every $8\%$ of the cumulative mass-loss rate.}.
            \label{fig:GasDist}
        \end{figure*}

        The gas density distribution in the inner disc after several hundred orbits is shown in Fig.~\ref{fig:GasDist} for the three simulated discs. After few hundred orbits at \SI{10}{au} the disc sets into a quasi-steady state in which the cavity inner edge has moved inwards from the initial location until an equilibrium between the viscous gas inflow and the photoevaporative gas removal is reached. This condition is different from the few tens of orbits needed to reach a vertically hydrostatic equilibrium, before dust insertion, as here the timescale is set by the concurrent viscous spreading and gas removal by the disc wind which take place over a longer timescale, depending on the alpha disc viscosity and the stellar mass and X-ray luminosity.

\begin{figure}
            \centering
            \includegraphics[width=\columnwidth]{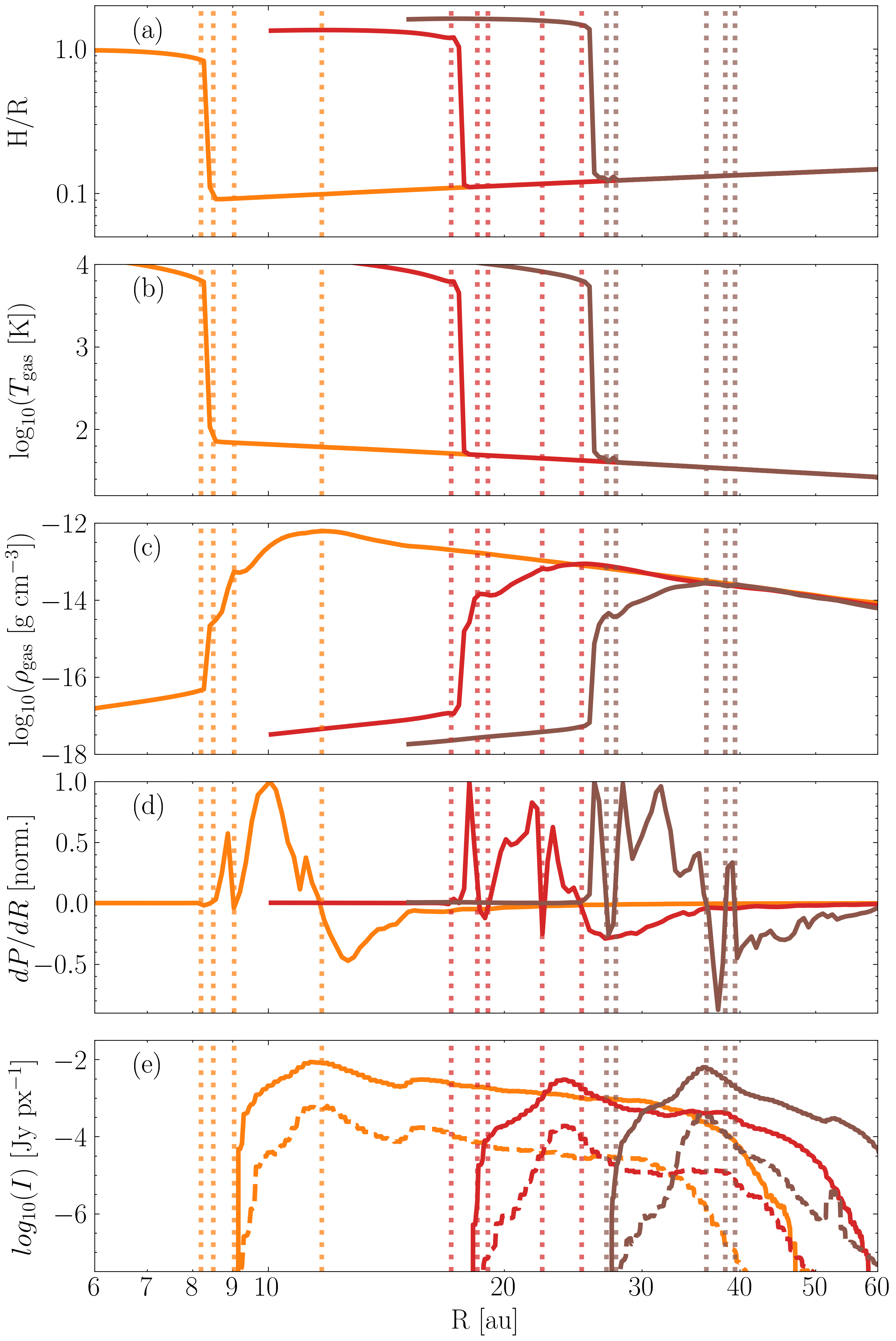}
            \caption{Panel a: disc scale height radial profile. Panel b: gas temperature radial profile. Panel c: mid-plane gas density radial profile. Panel d: normalized mid-plane pressure gradient. Panel e: radial profile of the continuum intensity in Band 3 (solid) and 7 (dashed dotted line). With dotted vertical lines the location of zero pressure gradient are reported. In all panels the orange, red, and brown lines represent the 10, 20, and 30 au cavity models respectively.
            \label{fig:scale-height}}
        \end{figure}

        The dashed lines mark the disc region which is directly affected by X-ray heating, while gas deeper in the disc is shielded from the star's radiation and in thermal equilibrium with the dust.
        The gas density is increased also close to the mid-plane in the cavity region. 
        This is due to the streamlines of the photoevaporative wind, which move radially inwards before carrying material outwards, as shown with red solid lines.
        At large radii the disc aspect ratio, flaring and gas density in the wind region are unaffected by the cavity size.
        Close to the cavity, as the unshielded X-ray irradiation heats up the bulk of the disc, the mid-plane temperature rises sharply and the disc scale height steeply increases as shown in Figure~\ref{fig:scale-height}a,b. Furthermore, plotting the pressure gradient at the disc mid-plane one can see that a double peak develops close to the inner gap edge (Figure~\ref{fig:scale-height}d).
        The innermost peak is created by the temperature increase, as the gas starts to be heated by the direct X-ray irradiation. The outermost one is related to the local maximum in the midplane gas density (see Figure~\ref{fig:scale-height}c).
        These effects have a strong impact on the gas and dust dynamics as shown in Section~\ref{sec:results-dust}, since the dust will collect at the locations where the pressure reaches a maximum, or the gradient is zero. In Figure~\ref{fig:scale-height}e, we show the intensity of the dust continuum emission at 0.85 and 3 mm for direct comparison.
        
    \subsection{Dust dynamics} \label{sec:results-dust}
\begin{figure*}
            \centering
            \includegraphics[width=\textwidth]{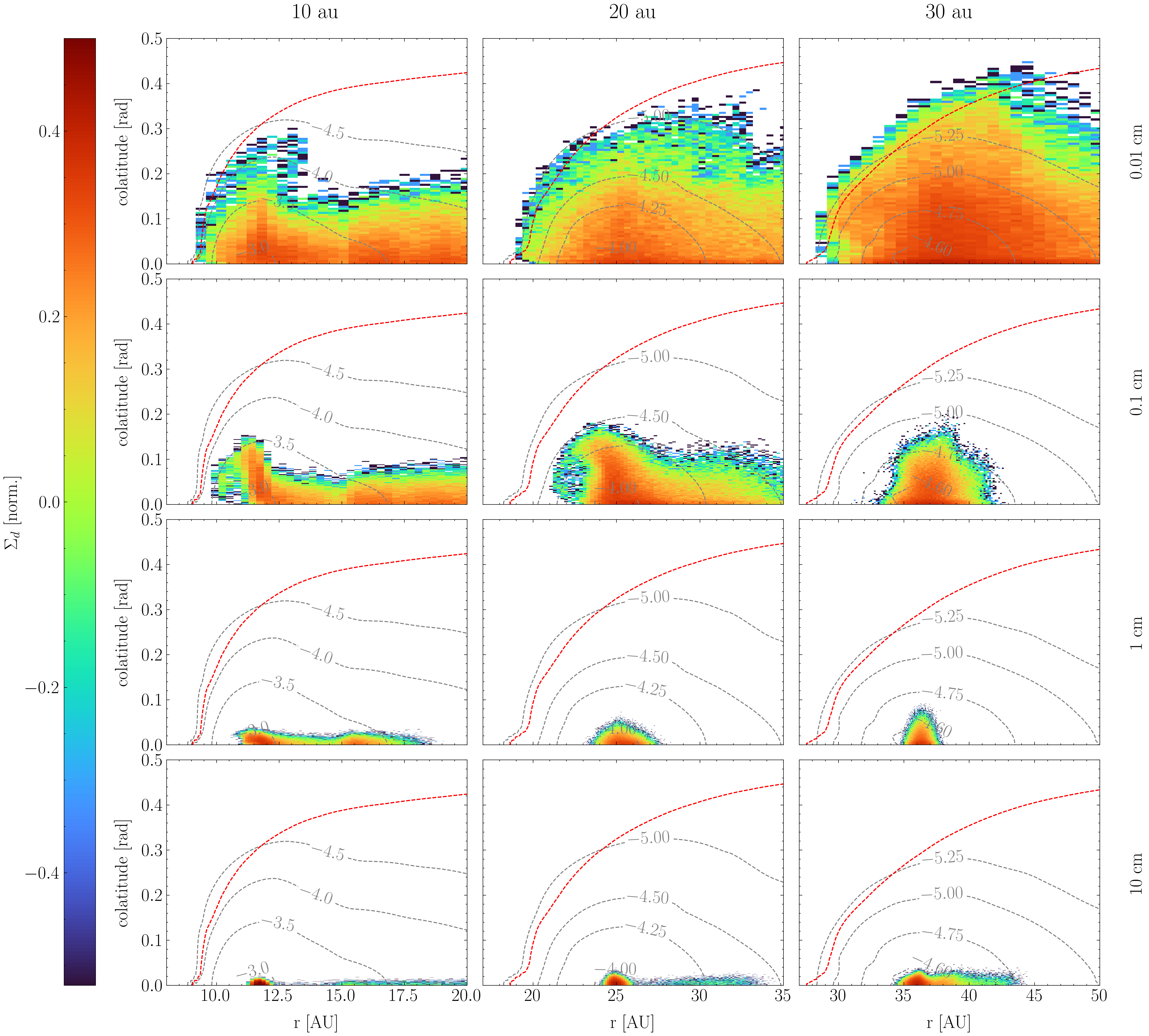}
            \caption{Dust (normalized) density distributions for the four dust particles modelled and the different cavity edges. The red dashed line show the maximum penetration depth of X-rays ($N_X = 2\cdot 10^{22}$ pp cm$^{-2}$), while the dashed black lines show the isocontours of gas pressure in log scale. The snapshot is taken after 1,500 yr of dust insertion.}
            \label{fig:GasDustDist}
        \end{figure*}

        After few tens of orbital periods of the gaseous disc, the dust component was included in the simulation.
        The strong vertical settling for particles with Stokes number close to unity (last two rows) is clearly visible in Figure~\ref{fig:GasDustDist} where the normalized dust surface densities are shown for the different runs over the underlying gas pressure profiles.
        In the radial direction the particles were initially spread over a range of \SI{30}{au} starting from the gas cavity. This means that the initial distribution reached out to \SI{40}{au} for the \SI{10}{au} cavity, \SI{50}{au} for the \SI{20}{au} cavity, and \SI{60}{au} for the \SI{30}{au} cavity.

        The smallest particles (St$\ll 1$) are very well coupled to the gas and thus follow its distribution closely, spreading to higher heights and very close to the inner cavity radius (see first row of Figure~\ref{fig:GasDustDist}).
        Particles of intermediate sizes ($\mathrm{St} \sim 1$) do not follow the gas distribution closely, though they interact strongly with the gas and migrate quickly towards the (second) gas pressure maxima (correspondent to the gas density maxima) just outside the gap.
        They also settle more efficiently following equation~\ref{eq:scale-height}.
        The effect of dust drift is evident comparing the four dust sizes, where the \SI{1}{cm}-sized particles (3rd row) have $\mathrm{St} \sim 1$ and are the ones drifting faster towards the pressure maxima at the gap edge.
        
        Furthermore, the particle scale height is strongly increased close to the inner cavity, since the gas temperature is higher due to the X-ray direct irradiation and the gas scale height is puffed up, as shown in Figure~\ref{fig:scale-height}a. This might allow a fraction of the smaller dust particles to be launched with the wind, however this effect is not visible in Figure~\ref{fig:GasDustDist}, since this fraction is too small to be captured by our limited number of modelled particles, and it involves mostly smaller (tightly coupled) dust particles. As explored recently in \citet{Franz2022}, the dust entrainment in transition discs can lead to potentially observable features that can be signatures of a thermal (or magnetic, see e.g. \citet{Rodenkirch2022}) wind.

        Interestingly, the particles with dust size equal to \SI{0.1}{cm} (second row) show a clear under-density at the disc mid-plane close to the cavity edge. This effect is cause by the vertical distribution of the pressure gradient, which is stronger at the disc midplane and gets weaker for increasing heights. The increased dust scale height close to the cavity edge allows particles of the same size to experience a large range of Stokes numbers. When they eventually settle towards the disc midplane their Stokes number increases and they drift directly towards the pressure maxima which deplets the region just inside it close to the disc midplane.
        
        The particle distribution at the end of the simulation suffers from the lack of constant dust flux from the outer edge due to our choice of a constant particle number (bottom rows of Figure~\ref{fig:GasDustDist}). As a result, for Stokes numbers close to unity, dust is mostly concentrated close to the disc cavity. In order to soften this effect we chose an intermediate time to post-process our results for these dust sizes since the time-scale to reach an equilibrium configuration is much faster for them.
  
    \subsection{Radiative Transfer} \label{sec:radiative-transfer}
\begin{figure*}
            \centering
            \includegraphics[width=\textwidth]{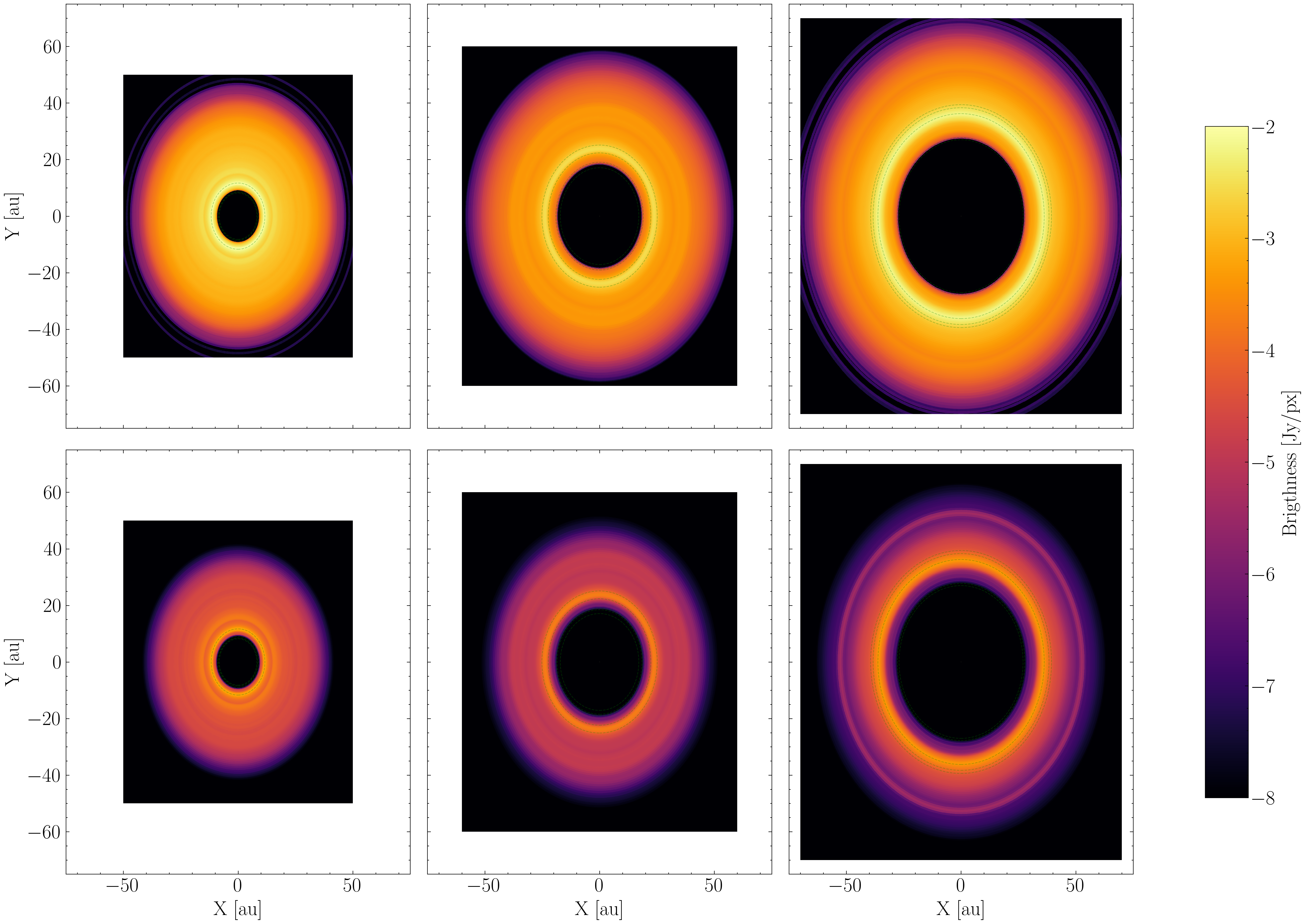}
            \caption{Continuum emission maps for the different transition discs cavities in Band 7 (top row) and Band 3 (bottom row). The location of the pressure maxima are highlighted with green dashed lines.}
            \label{fig:DustContinuum}
        \end{figure*}

        In Figure~\ref{fig:DustContinuum} we post-processed the equilibrium configuration of the dust distribution with \textsc{radmc-3d} obtaining the dust continuum emission at $0.85$ and \SI{3}{mm}, which represent Band 7 and 3 of ALMA, for the $10$, $20$ and \SI{30}{au} cases.
        
        The intensity of the dust emission has a maximum just beyond the gap edge, where the dust particles have accumulated at the location of the second pressure maxima where the gas density peaks (as highlighted with green dashed lines). This maximum takes the form of a bright ring, clearly visible in the \SI{3}{mm} continuum emission (see also Figure~\ref{fig:scale-height}e), as it collects more flux for particles with Stokes number close to unity.
        
        Beyond the bright ring the intensity gradually declines, as the dust at larger radii is increasingly shielded from stellar radiation. For the $20$ and \SI{30}{au} cases a second peak is visible in the outer disc, which corresponds to the flux of pebbles that is migrating from the outer edge and is not replenished by a constant flux.
        To allow for a more quantitative comparison of the emitted intensity we plot in Fig.~\ref{fig:comp_beam} (bottom row) a slice of the intensity map for the three discs at the two different wavelengths.
        The slices for the different wavelengths, which in principle trace different grain species, are of very similar shape and their maxima are located at the same radial distance, which doesn't allow us to have a quantitative observable of the effect of photoevaporation on different particle sizes (at least for the parameter space studied).
         
    \subsection{Synthetic Images} \label{sec:synthetic-images}
\begin{figure*}
          \centering
          \includegraphics[width=\textwidth]{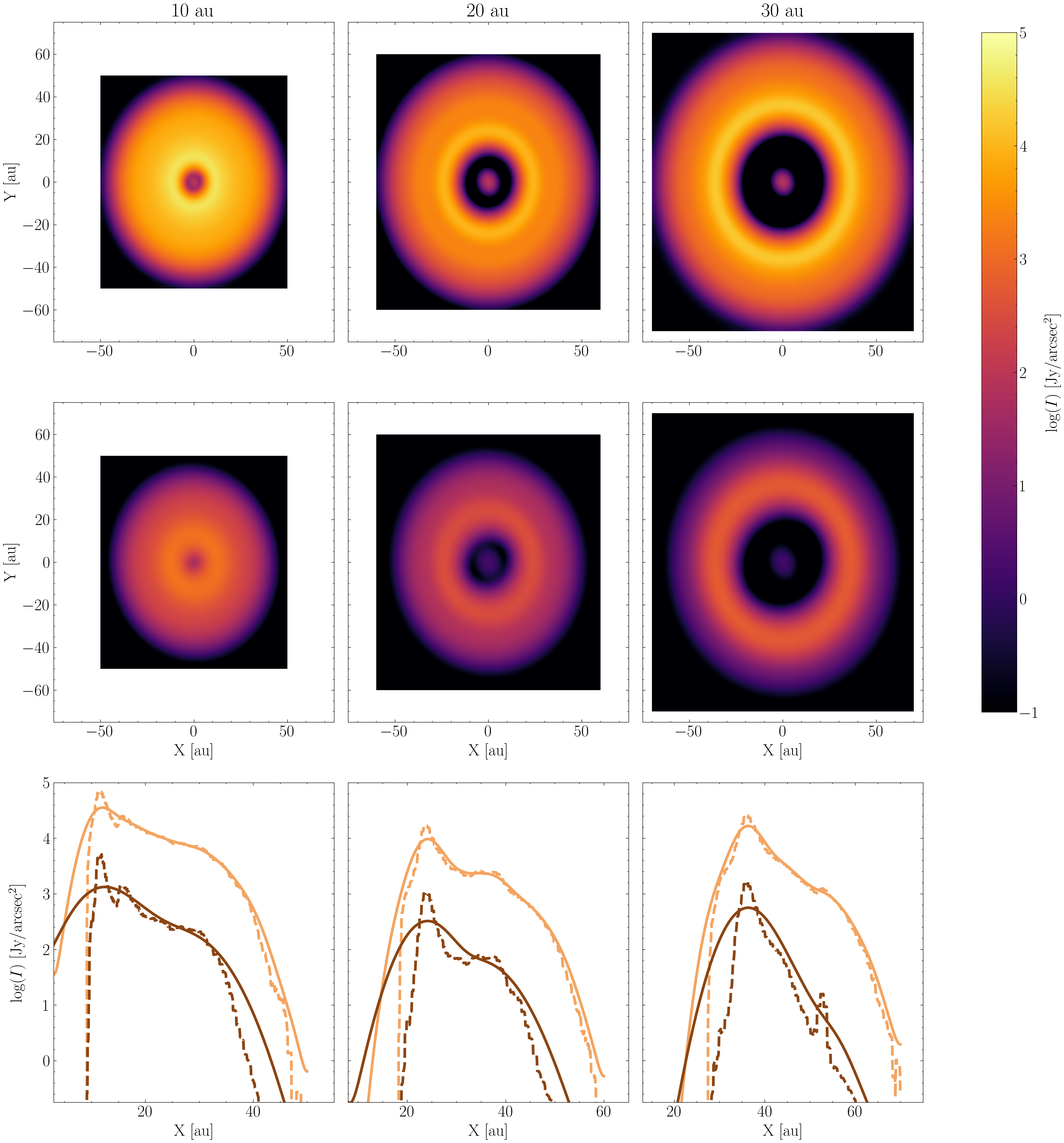}
          \caption{Dust continuum emission in Band 7 (top panel) and 3 (middle panel) convolved with a realistic ALMA beam size for the 10, 20, and 30 au transition discs. The convolution with the beam size blurs the features seen in Figure~\ref{fig:DustContinuum}. In the bottom panel the radial profiles are shown for a direct comparison between Band 7 (in orange) and Band 3 (in brown) and the convolved (solid lines) and unconvolved (dashed lines) intensities.}
          \label{fig:comp_beam}
        \end{figure*}

        We created mock observations by convolving the continuum dust emission with beam sizes of two ALMA configurations, as explained in Section~\ref{sec:radiative transfer}, in order to understand whether the observed features might be strong enough to be detected by current observational facilities and distinguishable from other physical processes. 
        The resulting images are shown in Fig.~\ref{fig:comp_beam}.
        In the cases of the \SI{3}{mm} and \SI{0.85}{mm} emission the clearly defined rings visible in Fig.~\ref{fig:GasDustDist} appear as broader features.
        The radial intensity curves for these mock observations are also shown in Fig.~\ref{fig:comp_beam} (bottom panel) and are smoothed out compared to the theoretical data but have their maximum at the same location, just outside the inner gap edge.
        
        Even though there are differences in the distributions of different particle sizes, the location of the intensity maximum is the same for \SI{0.85}{mm} and \SI{3}{mm} even in the theoretical images. 
        Despite the simplified way in which we simulated ALMA observations, 
        it is likely that the main rings in the \SI{0.85}{mm} and \SI{3}{mm} emission are bright and broad enough to be observable.
        
\section{Discussion} \label{sec:discussion}
    The features in the dust continuum emission formed by a photoevaporative wind should create a unique fingerprint that can be distinguished from other physical processes, given enough resolution.
    In order to measure in a quantitative way the difference between gap forming mechanisms we compared the intensity profiles of dust continuum emission for different physical processes, and the spectral index obtained from the two frequencies studied.

    \subsection{Dust continuum emission}
\begin{figure*}
        \centering
        \includegraphics[width=\textwidth]{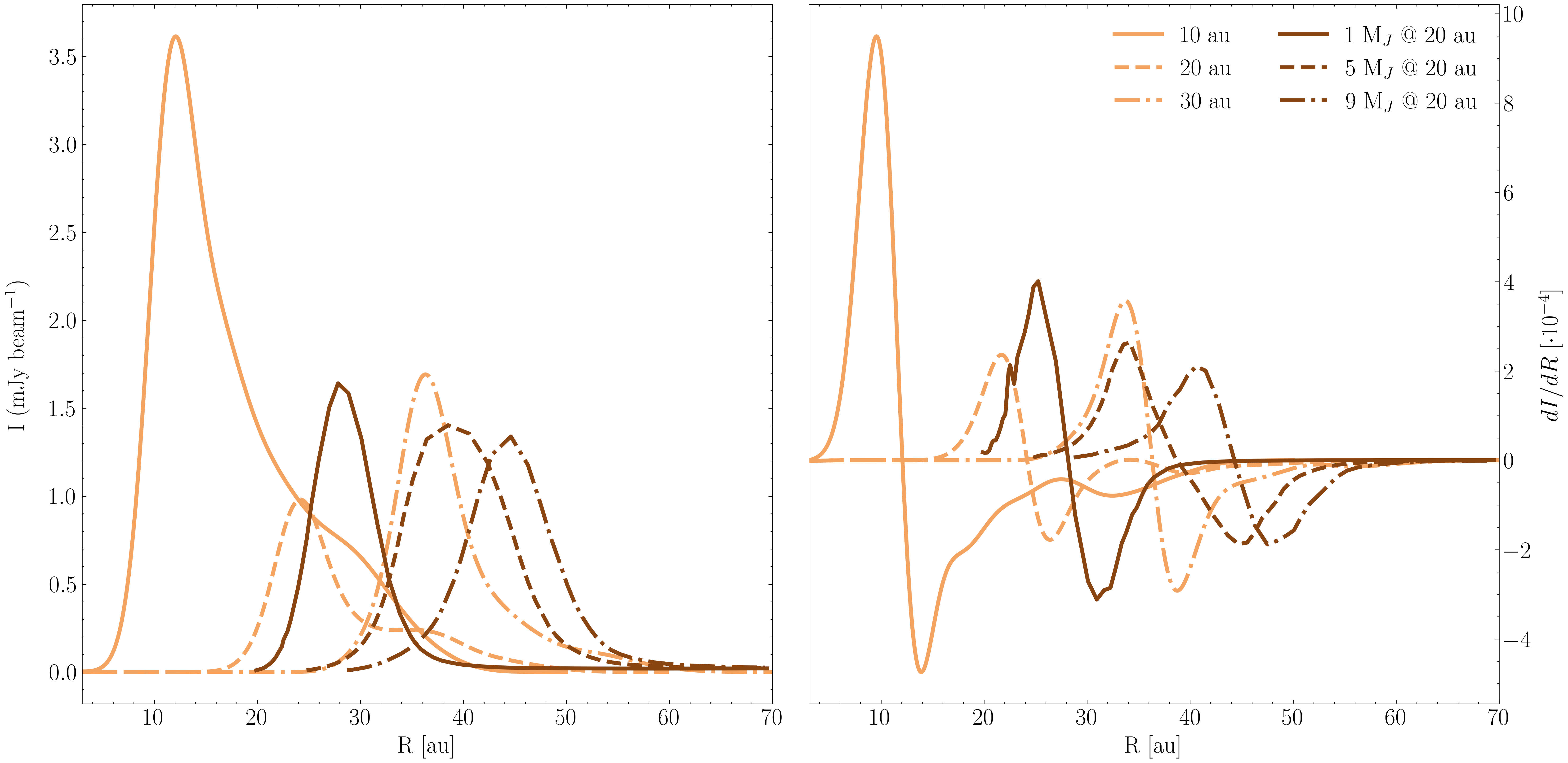}
        \caption{Panel a: radial intensity profile in Band 7 for this work in orange and from \citet{Ovelar2016,Facchini2018} for a $1$, $5$, and $9 M_\mathrm{J}$ mass planet in brown. Panel b: gradients of the intensity curves. We adopted a $0.03 \arcsec$ resolution beam and a distance of $150$ pc for direct comparison.}
        \label{fig:Grad}
    \end{figure*}

    The most straightforward way to create a gap is via planet-disc interaction. \citet{Facchini2018} modelled the dust continuum emission at \SI{0.85}{mm} wavelengths for a $1$, $5$, and \SI{9}{M_{Jup}} planets at \SI{20}{au} from the central star, postprocessing the hydrodynamical simulations by \cite{Ovelar2016}. We compared in Figure~\ref{fig:Grad} the intensity and radial gradient of the continuum emission in Band 7 (convolved with the same beam size adopted in their work, i.e. $0.03 \arcsec$ at a distance of \SI{150}{pc}) with our photoevaporative cavity. The intensity profile show a much stronger dependance for the transition discs as a function of the cavity radius with respect to the planetary masses at a fixed radius. On the one hand, the planetary case shows a clear trend of shallower profiles for increasing masses. On the other hand the photoevaporative gap shows a more complex behavior. Looking more in depth at the gradient of the profile in the right panel, we can see that the \SI{10}{au} case shows a much sharper gradient, given by the higher density close to the gap edge. The \SI{30}{au} case is comparable with a \SI{1}{M_\J} planet at \SI{20}{au} (dot dashed orange line and solid brown line) both looking at the intensity profile and its gradient. Potentially a \SI{1}{M_\J} at 27-\SI{28}{au} would generate a very similar intensity profile at the gap edge as a photoevaporative gap at \SI{30}{au}. The steepness of photoevaporative gaps seems then to increase as a function of gap cavity as the larger gap is able to collect an increasing function of the drifting dust. On the contrary a more massive giant planet will open a deeper and wider, but shallower gap for increasing masses. Thus, the only match between the two processes is for a 1 Jupiter mass planet, for the current choice of the parameter space (i.e. X-ray luminosity, disc properties).
    
    \subsection{Spectral index}
    To disentangle the planetary and photoevaporation scenario one can extract from a multi-wavelength analysis of an object its spectral index, defined as
    \begin{equation}
        \alpha_{73} = \frac{\log{(I_7/I_3)}}{\log{(\nu_7/\nu_3)}}\,,
    \end{equation}
    where $I_7$ and $\nu_7$ are the intensity and frequency in Band 7, and $I_3$, $\nu_3$ are the intensity and frequency in Band 3.
    The spectral index gives a measure of dust growth or dust dynamical configuration, in the optical thin approximation.
    \begin{figure}
        \centering
        \includegraphics[width=\columnwidth]{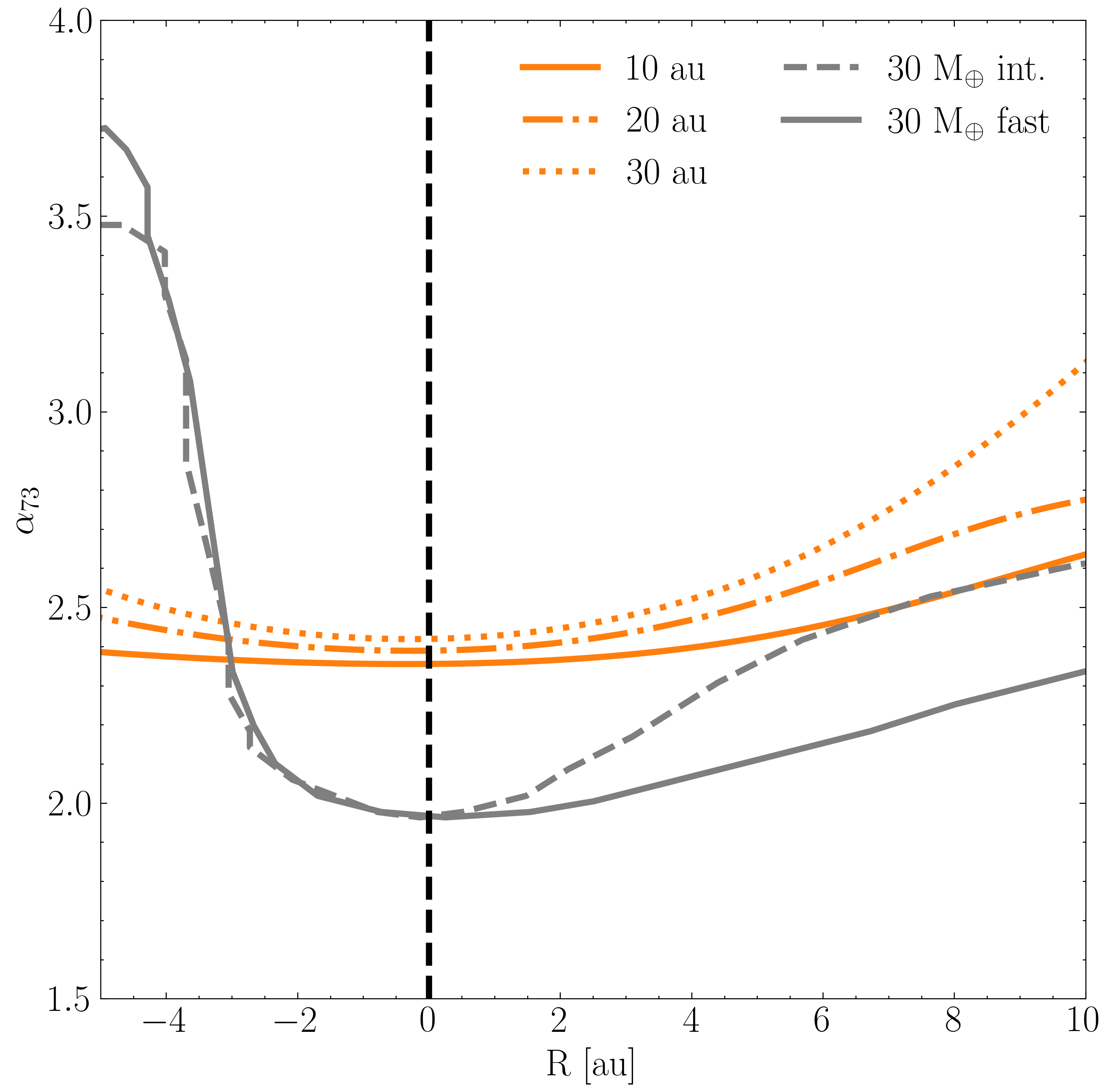}
        \caption{Spectral indices calculated from the synthetic images in Band 3 and 7, for the three different transition disc models, compared with the intermediate and fast migration case from \citet{Nazari_2019}, shifted to be centred at the cavity edge and zoomed in the cavity region.}
        \label{fig:spectral_index}
    \end{figure}
    In Figure~\ref{fig:spectral_index} we plotted the spectral index for our three transition disc simulations, where the profiles have been shifted to give a direct comparison and zoomed in close to the cavity edge, that has been highlighted with a vertical dashed black line.
    The two main trends that are observed are a reduction of the spectral index in the bright ring corresponding to the gap edge of the transition disc, and a general increase of the spectral index as a function of the distance from the central star, which is compatible with the current observations \citep[e.g.][]{Long2020}.
    The most notable difference between our models is that for smaller cavities there is a less steep increase of the spectral index, since large dust particles are not all collected close to the inner edge of the cavity but spread across a larger region.
    Since we do not include dust growth in these simulations, this is purely a dynamical effect. The reason for this difference is the limited dust supply from the outer disc region, and the larger Stokes number for dust particles of the same size in the outer disc (because of the decreased gas surface density, see eq.~\ref{eq:Stokes_number}) which increases their drift speed towards the cavity edge.

    We compared our results with \citet{Nazari_2019}, that modelled the observability of a \SI{30}{M_\oplus} planet with different orbital migration for the same bands.
    They assumed a different approach than in the current work, modelling the planet-disc interaction in 2D (assuming a vertical dust distribution a posteriori), and adopting a pressureless fluid approach to describe the dust evolution, which does not allow to follow the history of a particular dust sample but lessen the problem of a limited dust supply to the cavity edge.
    There is a stark difference with respect to the planetary case. 
    The minimum spectral index at the ring location is larger for the photoevaporative case which implies a less strong filtering effect with respect to the \SI{30}{M_\oplus} planet. 
    Inside the cavity the spectral index increases again due to the low density and low optical depth of the disc wind.

    There is a growing literature of multi-wavelength high-resolution observations of planet forming discs, of which few of them cover also transition and pre-transition discs. 
    In particuar, the HD169142 system \citep{Macias2019} shows two bright rings at $25$ and \SI{60}{au} with spectral index between Band $3$ and $7$ of $2.025\pm0.03$ and $2.33\pm0.03$ respectively, where the first is only compatible with a planetary origin while the second is closer to the predictions of disk photoevaporation. Indeed it has been recently suggested that a Jupiter mass planet might be hiding in the disc cavity \citep{Hammond2023}, which render this system an important test bed in order to constraint current modelling.
    Furthermore, the transition disc GM Aur \citep{Huang2020} shows two bright rings at 40 and \SI{84}{au} with a spectral index between 1.1 and 2.1 mm of 2.1 and 2.35, respectively. Here again the first value is compatible with the planetary scenario, while the second one is on the high end, but a targeted analysis of these systems should be made to draw meaningful conclusions, which goes beyond the scope of the current work.

    \subsection{Model limitations}

    The spectral index comparison with a \SI{30}{M_\oplus} planet is inconsistent with the more massive ones previously considered for the dust continuum emission, though a jupiter mass planet will create a steeper cavity which will have a stronger filtering effect, enhanching the behavior observed for smaller mass planerts.
    However, recent results point to more leaky giant planets \citep[see e.g.][]{Stammler2023} which might alleviate the observed difference in the spectral indeces between these two effects.
    Nevertheless, a self-consistent modelling of planet and photoevaporation within the same framework would help better to disentangle the two effects, and it is indeed our plan for the future. Further differences given by the non axysimmetric structures introduced by the planet, and the temporal variability on an orbital timescale can be used to further distinguish a planetary induced transition disc.

    In this study we considered only a very specific set of parameters. In the future we plan to relax these conditions by varying the disc (alpha, disc mass) and stellar properties (stellar mass and X-ray luminosities). Furthermore, to expand the parameter space, a good avenue would be to restrict to a 1D disc evolution with prescribed wind mass-loss rates that would allow us to include dust growth/fragmentation, similarly as in \citet{Garate2021}.

\section{Conclusions} \label{sec:conclusions}

We modelled the dust dynamics in a 2D dusty hydrodynamical simulation of a transition disc carved by stellar X(EUV) irradiation from the central star, and found the equilibrium distribution of a range of dust sizes.
From this result we calculated the dust continuum emission at $0.85$ and \SI{3}{mm} wavelengths and derived synthetic ALMA observations. We then compared the cavity profiles created by photoevaporation with those of transition discs carved by a super-Earth/giant planets, in order to find characteristic features of each physical process that can be observed in current and future observational campaigns.
We found that the steepness of the 30 au photoevaporated dust cavity is similar to that of a 1 M$_\mathrm{J}$ planet (at a similar location), while larger planets and/or larger photoevaporated cavities can generates distinguishable features.
In order to further disentangle these two effects, the spectral index analysis shows very different profiles for the planetary and photoevaporative cases, with a higher spectral index at the photoevaporated cavity edge due to the lower filtering efficiency.
This combination of features might be specific to transition discs opened by photoevaporation, though a larger study for different stellar X-ray luminosity and stellar mass, as well as a direct comparison to a transition disc with similar physical properties opened by a planet should be necessary to better constrain the parameter space covered by stellar photoevaporation.

\section*{Acknowledgements}
We thank the anonymous referees for the useful comments and suggestions that substantially helped improving the quality of the paper.
This research was supported by the Excellence Cluster ORIGINS which is funded by the Deutsche Forschungsgemeinschaft (DFG, German Research Foundation) under Germany's Excellence Strategy - EXC-2094-390783311. BE and GP acknowledge the support of the Deutsche Forschungsgemeinschaft (DFG, German Research Foundation) - 325594231.
This work was performed on the computing facilities of the Computational Center for Particle and Astrophysics (C2PAP).
This paper utilizes the D'Alessio Irradiated Accretion Disk (\textsc{diad}) code. We wish to recognize the work of Paola D'Alessio, who passed away in 2013. Her legacy and pioneering work live on through her substantial contributions to the field.

\section*{Data Availability}
The data underlying this article and the scripts used to create the Figures are available at \href{https://github.com/GiovanniPicogna/observability-td}{https://github.com/GiovanniPicogna/observability-td}.


\bibliographystyle{mnras}
\bibliography{lib}

\begin{thebibliography}{}
\makeatletter
\relax
\def\mn@urlcharsother{\let\do\@makeother \do\$\do\&\do\#\do\^\do\_\do\%\do\~}
\def\mn@doi{\begingroup\mn@urlcharsother \@ifnextchar [ {\mn@doi@}
  {\mn@doi@[]}}
\def\mn@doi@[#1]#2{\def\@tempa{#1}\ifx\@tempa\@empty \href
  {http://dx.doi.org/#2} {doi:#2}\else \href {http://dx.doi.org/#2} {#1}\fi
  \endgroup}
\def\mn@eprint#1#2{\mn@eprint@#1:#2::\@nil}
\def\mn@eprint@arXiv#1{\href {http://arxiv.org/abs/#1} {{\tt arXiv:#1}}}
\def\mn@eprint@dblp#1{\href {http://dblp.uni-trier.de/rec/bibtex/#1.xml}
  {dblp:#1}}
\def\mn@eprint@#1:#2:#3:#4\@nil{\def\@tempa {#1}\def\@tempb {#2}\def\@tempc
  {#3}\ifx \@tempc \@empty \let \@tempc \@tempb \let \@tempb \@tempa \fi \ifx
  \@tempb \@empty \def\@tempb {arXiv}\fi \@ifundefined
  {mn@eprint@\@tempb}{\@tempb:\@tempc}{\expandafter \expandafter \csname
  mn@eprint@\@tempb\endcsname \expandafter{\@tempc}}}

\bibitem[\protect\citeauthoryear{{Alexander}, {Pascucci}, {Andrews}, {Armitage}
   \& {Cieza}}{{Alexander} et~al.}{2014}]{Alexander2014}
{Alexander} R.,  {Pascucci} I.,  {Andrews} S.,  {Armitage} P.,   {Cieza} L.,
  2014, in {Beuther} H.,  {Klessen} R.~S.,  {Dullemond} C.~P.,   {Henning} T.,
  eds, Protostars and Planets VI. p.~475 (\mn@eprint {arXiv} {1311.1819}),
  \mn@doi{10.2458/azu\_uapress\_9780816531240-ch021}

\bibitem[\protect\citeauthoryear{{Andrews} et~al.,}{{Andrews}
  et~al.}{2018}]{Andrews2018}
{Andrews} S.~M.,  et~al., 2018, \mn@doi [\apjl] {10.3847/2041-8213/aaf741},
  \href {https://ui.adsabs.harvard.edu/abs/2018ApJ...869L..41A} {869, L41}

\bibitem[\protect\citeauthoryear{{Baruteau} et~al.,}{{Baruteau}
  et~al.}{2019}]{Baruteau2019}
{Baruteau} C.,  et~al., 2019, \mn@doi [\mnras] {10.1093/mnras/stz802}, \href
  {https://ui.adsabs.harvard.edu/abs/2019MNRAS.486..304B} {486, 304}

\bibitem[\protect\citeauthoryear{Clarke, Gendrin  \& Sotomayor}{Clarke
  et~al.}{2001}]{Clarke2001}
Clarke C.~J.,  Gendrin A.,   Sotomayor M.,  2001, \mn@doi [Monthly Notices of
  the Royal Astronomical Society] {10.1046/j.1365-8711.2001.04891.x}, 328, 485

\bibitem[\protect\citeauthoryear{{D'Alessio}, {Cant{\"o}}, {Calvet}  \&
  {Lizano}}{{D'Alessio} et~al.}{1998}]{DAlessio1998}
{D'Alessio} P.,  {Cant{\"o}} J.,  {Calvet} N.,   {Lizano} S.,  1998, \mn@doi
  [\apj] {10.1086/305702}, \href
  {https://ui.adsabs.harvard.edu/abs/1998ApJ...500..411D} {500, 411}

\bibitem[\protect\citeauthoryear{{D'Alessio}, {Calvet}, {Hartmann}, {Lizano}
  \& {Cant{\'o}}}{{D'Alessio} et~al.}{1999}]{DAlessio1999}
{D'Alessio} P.,  {Calvet} N.,  {Hartmann} L.,  {Lizano} S.,   {Cant{\'o}} J.,
  1999, \mn@doi [\apj] {10.1086/308103}, \href
  {https://ui.adsabs.harvard.edu/abs/1999ApJ...527..893D} {527, 893}

\bibitem[\protect\citeauthoryear{{D'Alessio}, {Calvet}  \&
  {Hartmann}}{{D'Alessio} et~al.}{2001}]{DAlessio2001}
{D'Alessio} P.,  {Calvet} N.,   {Hartmann} L.,  2001, \mn@doi [\apj]
  {10.1086/320655}, \href
  {https://ui.adsabs.harvard.edu/abs/2001ApJ...553..321D} {553, 321}

\bibitem[\protect\citeauthoryear{{D'Alessio} et~al.,}{{D'Alessio}
  et~al.}{2005}]{DAlessio2005}
{D'Alessio} P.,  et~al., 2005, \mn@doi [\apj] {10.1086/427490}, \href
  {https://ui.adsabs.harvard.edu/abs/2005ApJ...621..461D} {621, 461}

\bibitem[\protect\citeauthoryear{{D'Alessio}, {Calvet}, {Hartmann},
  {Franco-Hern{\'a}ndez}  \& {Serv{\'\i}n}}{{D'Alessio}
  et~al.}{2006}]{DAlessio2006}
{D'Alessio} P.,  {Calvet} N.,  {Hartmann} L.,  {Franco-Hern{\'a}ndez} R.,
  {Serv{\'\i}n} H.,  2006, \mn@doi [\apj] {10.1086/498861}, \href
  {https://ui.adsabs.harvard.edu/abs/2006ApJ...638..314D} {638, 314}

\bibitem[\protect\citeauthoryear{{Dong} \& {Dawson}}{{Dong} \&
  {Dawson}}{2016}]{Dong2016}
{Dong} R.,  {Dawson} R.,  2016, \mn@doi [\apj] {10.3847/0004-637X/825/1/77},
  \href {https://ui.adsabs.harvard.edu/abs/2016ApJ...825...77D} {825, 77}

\bibitem[\protect\citeauthoryear{{Draine}}{{Draine}}{2003}]{Draine2003}
{Draine} B.~T.,  2003, \mn@doi [\apj] {10.1086/379123}, \href
  {https://ui.adsabs.harvard.edu/abs/2003ApJ...598.1026D} {598, 1026}

\bibitem[\protect\citeauthoryear{{Duffell} \& {Dong}}{{Duffell} \&
  {Dong}}{2015}]{Duffell2015}
{Duffell} P.~C.,  {Dong} R.,  2015, \mn@doi [\apj]
  {10.1088/0004-637X/802/1/42}, \href
  {https://ui.adsabs.harvard.edu/abs/2015ApJ...802...42D} {802, 42}

\bibitem[\protect\citeauthoryear{{Dullemond}, {Juhasz}, {Pohl}, {Sereshti},
  {Shetty}, {Peters}, {Commercon}  \& {Flock}}{{Dullemond}
  et~al.}{2012}]{radmc3d}
{Dullemond} C.~P.,  {Juhasz} A.,  {Pohl} A.,  {Sereshti} F.,  {Shetty} R.,
  {Peters} T.,  {Commercon} B.,   {Flock} M.,  2012, {RADMC-3D: A multi-purpose
  radiative transfer tool} (\mn@eprint {ascl} {1202.015})

\bibitem[\protect\citeauthoryear{{Ercolano} \& {Pascucci}}{{Ercolano} \&
  {Pascucci}}{2017}]{Ercolano2017}
{Ercolano} B.,  {Pascucci} I.,  2017, \mn@doi [Royal Society Open Science]
  {10.1098/rsos.170114}, \href
  {https://ui.adsabs.harvard.edu/abs/2017RSOS....470114E} {4, 170114}

\bibitem[\protect\citeauthoryear{{Ercolano}, {Barlow}, {Storey}  \&
  {Liu}}{{Ercolano} et~al.}{2003}]{MOCASSIN1}
{Ercolano} B.,  {Barlow} M.~J.,  {Storey} P.~J.,   {Liu} X.-W.,  2003, \mn@doi
  [\mnras] {10.1046/j.1365-8711.2003.06371.x}, \href
  {http://adsabs.harvard.edu/abs/2003MNRAS.340.1136E} {340, 1136}

\bibitem[\protect\citeauthoryear{{Ercolano}, {Barlow}  \& {Storey}}{{Ercolano}
  et~al.}{2005}]{MOCASSIN2}
{Ercolano} B.,  {Barlow} M.~J.,   {Storey} P.~J.,  2005, \mn@doi [\mnras]
  {10.1111/j.1365-2966.2005.09381.x}, \href
  {http://adsabs.harvard.edu/abs/2005MNRAS.362.1038E} {362, 1038}

\bibitem[\protect\citeauthoryear{{Ercolano}, {Young}, {Drake}  \&
  {Raymond}}{{Ercolano} et~al.}{2008a}]{MOCASSIN3}
{Ercolano} B.,  {Young} P.~R.,  {Drake} J.~J.,   {Raymond} J.~C.,  2008a,
  \mn@doi [\apjs] {10.1086/524378}, \href
  {http://adsabs.harvard.edu/abs/2008ApJS..175..534E} {175, 534}

\bibitem[\protect\citeauthoryear{{Ercolano}, {Drake}, {Raymond}  \&
  {Clarke}}{{Ercolano} et~al.}{2008b}]{Ercolano2008}
{Ercolano} B.,  {Drake} J.~J.,  {Raymond} J.~C.,   {Clarke} C.~C.,  2008b,
  \mn@doi [\apj] {10.1086/590490}, \href
  {http://adsabs.harvard.edu/abs/2008ApJ...688..398E} {688, 398}

\bibitem[\protect\citeauthoryear{{Ercolano}, {Clarke}  \& {Drake}}{{Ercolano}
  et~al.}{2009}]{Ercolano2009}
{Ercolano} B.,  {Clarke} C.~J.,   {Drake} J.~J.,  2009, \mn@doi [\apj]
  {10.1088/0004-637X/699/2/1639}, \href
  {http://adsabs.harvard.edu/abs/2009ApJ...699.1639E} {699, 1639}

\bibitem[\protect\citeauthoryear{{Ercolano}, {Weber}  \& {Owen}}{{Ercolano}
  et~al.}{2018}]{Ercolano2018}
{Ercolano} B.,  {Weber} M.~L.,   {Owen} J.~E.,  2018, \mn@doi [\mnras]
  {10.1093/mnrasl/slx168}, \href
  {https://ui.adsabs.harvard.edu/abs/2018MNRAS.473L..64E} {473, L64}

\bibitem[\protect\citeauthoryear{{Facchini}, {Pinilla}, {van Dishoeck}  \& {de
  Juan Ovelar}}{{Facchini} et~al.}{2018}]{Facchini2018}
{Facchini} S.,  {Pinilla} P.,  {van Dishoeck} E.~F.,   {de Juan Ovelar} M.,
  2018, \mn@doi [\aap] {10.1051/0004-6361/201731390}, \href
  {https://ui.adsabs.harvard.edu/abs/2018A&A...612A.104F} {612, A104}

\bibitem[\protect\citeauthoryear{{Franz}, {Picogna}, {Ercolano}, {Casassus},
  {Birnstiel}, {Rab}  \& {P{\'e}rez}}{{Franz} et~al.}{2022}]{Franz2022}
{Franz} R.,  {Picogna} G.,  {Ercolano} B.,  {Casassus} S.,  {Birnstiel} T.,
  {Rab} C.,   {P{\'e}rez} S.,  2022, \mn@doi [\aap]
  {10.1051/0004-6361/202142785}, \href
  {https://ui.adsabs.harvard.edu/abs/2022A&A...659A..90F} {659, A90}

\bibitem[\protect\citeauthoryear{{G{\'a}rate} et~al.,}{{G{\'a}rate}
  et~al.}{2021}]{Garate2021}
{G{\'a}rate} M.,  et~al., 2021, \mn@doi [\aap] {10.1051/0004-6361/202141444},
  \href {https://ui.adsabs.harvard.edu/abs/2021A&A...655A..18G} {655, A18}

\bibitem[\protect\citeauthoryear{{Hammond}, {Christiaens}, {Price}, {Toci},
  {Pinte}, {Juillard}  \& {Garg}}{{Hammond} et~al.}{2023}]{Hammond2023}
{Hammond} I.,  {Christiaens} V.,  {Price} D.~J.,  {Toci} C.,  {Pinte} C.,
  {Juillard} S.,   {Garg} H.,  2023, \mn@doi [\mnras] {10.1093/mnrasl/slad027},
  \href {https://ui.adsabs.harvard.edu/abs/2023MNRAS.tmpL..30H} {}

\bibitem[\protect\citeauthoryear{{Hollenbach}, {Johnstone}, {Lizano}  \&
  {Shu}}{{Hollenbach} et~al.}{1994}]{Hollenbach1994}
{Hollenbach} D.,  {Johnstone} D.,  {Lizano} S.,   {Shu} F.,  1994, \mn@doi
  [\apj] {10.1086/174276}, \href
  {https://ui.adsabs.harvard.edu/abs/1994ApJ...428..654H} {428, 654}

\bibitem[\protect\citeauthoryear{{Huang} et~al.,}{{Huang}
  et~al.}{2020}]{Huang2020}
{Huang} J.,  et~al., 2020, \mn@doi [\apj] {10.3847/1538-4357/ab711e}, \href
  {https://ui.adsabs.harvard.edu/abs/2020ApJ...891...48H} {891, 48}

\bibitem[\protect\citeauthoryear{{Kurtovic} et~al.,}{{Kurtovic}
  et~al.}{2021}]{Kurtovic2021}
{Kurtovic} N.~T.,  et~al., 2021, \mn@doi [\aap] {10.1051/0004-6361/202038983},
  \href {https://ui.adsabs.harvard.edu/abs/2021A&A...645A.139K} {645, A139}

\bibitem[\protect\citeauthoryear{{Long} et~al.,}{{Long}
  et~al.}{2020}]{Long2020}
{Long} F.,  et~al., 2020, \mn@doi [\apj] {10.3847/1538-4357/ab9a54}, \href
  {https://ui.adsabs.harvard.edu/abs/2020ApJ...898...36L} {898, 36}

\bibitem[\protect\citeauthoryear{{Mac{\'\i}as} et~al.,}{{Mac{\'\i}as}
  et~al.}{2019}]{Macias2019}
{Mac{\'\i}as} E.,  et~al., 2019, \mn@doi [\apj] {10.3847/1538-4357/ab31a2},
  \href {https://ui.adsabs.harvard.edu/abs/2019ApJ...881..159M} {881, 159}

\bibitem[\protect\citeauthoryear{{Mignone}, {Bodo}, {Massaglia}, {Matsakos},
  {Tesileanu}, {Zanni}  \& {Ferrari}}{{Mignone} et~al.}{2007}]{Mignone2007}
{Mignone} A.,  {Bodo} G.,  {Massaglia} S.,  {Matsakos} T.,  {Tesileanu} O.,
  {Zanni} C.,   {Ferrari} A.,  2007, \mn@doi [\apjs] {10.1086/513316}, \href
  {http://adsabs.harvard.edu/abs/2007ApJS..170..228M} {170, 228}

\bibitem[\protect\citeauthoryear{Nazari, Booth, Clarke, Rosotti, Tazzari,
  Juhasz  \& Meru}{Nazari et~al.}{2019}]{Nazari_2019}
Nazari P.,  Booth R.~A.,  Clarke C.~J.,  Rosotti G.~P.,  Tazzari M.,  Juhasz
  A.,   Meru F.,  2019, \mn@doi [Monthly Notices of the Royal Astronomical
  Society] {10.1093/mnras/stz836}, 485, 5914–5923

\bibitem[\protect\citeauthoryear{{Nielsen} et~al.,}{{Nielsen}
  et~al.}{2019}]{Nielsen2019}
{Nielsen} E.~L.,  et~al., 2019, \mn@doi [\aj] {10.3847/1538-3881/ab16e9}, \href
  {https://ui.adsabs.harvard.edu/abs/2019AJ....158...13N} {158, 13}

\bibitem[\protect\citeauthoryear{{Picogna}, {Stoll}  \& {Kley}}{{Picogna}
  et~al.}{2018}]{Picogna2018}
{Picogna} G.,  {Stoll} M. H.~R.,   {Kley} W.,  2018, \mn@doi [\aap]
  {10.1051/0004-6361/201732523}, \href
  {https://ui.adsabs.harvard.edu/abs/2018A&A...616A.116P} {616, A116}

\bibitem[\protect\citeauthoryear{{Picogna}, {Ercolano}, {Owen}  \&
  {Weber}}{{Picogna} et~al.}{2019}]{Picogna2019}
{Picogna} G.,  {Ercolano} B.,  {Owen} J.~E.,   {Weber} M.~L.,  2019, \mn@doi
  [\mnras] {10.1093/mnras/stz1166}, \href
  {https://ui.adsabs.harvard.edu/abs/2019MNRAS.487..691P} {487, 691}

\bibitem[\protect\citeauthoryear{{Rodenkirch} \& {Dullemond}}{{Rodenkirch} \&
  {Dullemond}}{2022}]{Rodenkirch2022}
{Rodenkirch} P.~J.,  {Dullemond} C.~P.,  2022, \mn@doi [\aap]
  {10.1051/0004-6361/202142571}, \href
  {https://ui.adsabs.harvard.edu/abs/2022A&A...659A..42R} {659, A42}

\bibitem[\protect\citeauthoryear{{Stammler}, {Lichtenberg},
  {Dr{\k{a}}{\.z}kowska}  \& {Birnstiel}}{{Stammler}
  et~al.}{2023}]{Stammler2023}
{Stammler} S.~M.,  {Lichtenberg} T.,  {Dr{\k{a}}{\.z}kowska} J.,   {Birnstiel}
  T.,  2023, \mn@doi [\aap] {10.1051/0004-6361/202245512}, \href
  {https://ui.adsabs.harvard.edu/abs/2023A&A...670L...5S} {670, L5}

\bibitem[\protect\citeauthoryear{{Strom}, {Strom}, {Edwards}, {Cabrit}  \&
  {Skrutskie}}{{Strom} et~al.}{1989}]{Strom1989}
{Strom} K.~M.,  {Strom} S.~E.,  {Edwards} S.,  {Cabrit} S.,   {Skrutskie}
  M.~F.,  1989, \mn@doi [\aj] {10.1086/115085}, \href
  {https://ui.adsabs.harvard.edu/abs/1989AJ.....97.1451S} {97, 1451}

\bibitem[\protect\citeauthoryear{{Vigan} et~al.,}{{Vigan}
  et~al.}{2021}]{Vigan2021}
{Vigan} A.,  et~al., 2021, \mn@doi [\aap] {10.1051/0004-6361/202038107}, \href
  {https://ui.adsabs.harvard.edu/abs/2021A&A...651A..72V} {651, A72}

\bibitem[\protect\citeauthoryear{{Weidenschilling}}{{Weidenschilling}}{1977}]{Weidenschilling1977}
{Weidenschilling} S.~J.,  1977, \mn@doi [\mnras] {10.1093/mnras/180.2.57},
  \href {https://ui.adsabs.harvard.edu/abs/1977MNRAS.180...57W} {180, 57}

\bibitem[\protect\citeauthoryear{{Whipple}}{{Whipple}}{1972}]{Whipple1972}
{Whipple} F.~L.,  1972, in {Elvius} A.,  ed., From Plasma to Planet. p.~211

\bibitem[\protect\citeauthoryear{{Youdin} \& {Lithwick}}{{Youdin} \&
  {Lithwick}}{2007}]{Youdin2007}
{Youdin} A.~N.,  {Lithwick} Y.,  2007, \mn@doi [\icarus]
  {10.1016/j.icarus.2007.07.012}, \href
  {https://ui.adsabs.harvard.edu/abs/2007Icar..192..588Y} {192, 588}

\bibitem[\protect\citeauthoryear{{de Juan Ovelar}, {Pinilla}, {Min}, {Dominik}
  \& {Birnstiel}}{{de Juan Ovelar} et~al.}{2016}]{Ovelar2016}
{de Juan Ovelar} M.,  {Pinilla} P.,  {Min} M.,  {Dominik} C.,   {Birnstiel} T.,
   2016, \mn@doi [\mnras] {10.1093/mnrasl/slw051}, \href
  {https://ui.adsabs.harvard.edu/abs/2016MNRAS.459L..85D} {459, L85}

\bibitem[\protect\citeauthoryear{{van der Marel} et~al.,}{{van der Marel}
  et~al.}{2022}]{vanderMarel2022}
{van der Marel} N.,  et~al., 2022, arXiv e-prints, \href
  {https://ui.adsabs.harvard.edu/abs/2022arXiv220408225V} {p. arXiv:2204.08225}

\makeatother
\end{thebibliography}





\bsp	
\label{lastpage}
\end{document}